\documentstyle[prd,aps]{revtex}
\tighten
\begin{document}
\draft

\title{On the interpretation of time-reparametrization-invariant
quantum mechanics}

\author{Ian D. Lawrie}
\address{Department of Physics, University of Leeds, Leeds LS2 9JT,
UK}

\author{Richard J. Epp}
\address{Department of Physics, University of Winnipeg, Winnipeg,
Manitoba, Canada}
\address{Department of Physics, University of California, Davis CA
95616, USA\thanks{Present address} }

\maketitle

\begin{abstract}
The classical and quantum dynamics of simple time-reparametrization-
invariant models containing two degrees of freedom are studied in
detail. Elimination of one ``clock'' variable through the Hamiltonian
constraint leads to a description of time evolution for the remaining
variable which is essentially equivalent to the standard quantum
mechanics of an unconstrained system.  In contrast to a similar
proposal of Rovelli, evolution is with respect to the geometrical
proper time, and the Heisenberg equation of motion is exact.  The
possibility of a ``test clock'', which would reveal time evolution
while contributing negligibly to the Hamiltonian constraint is
examined, and found to be viable in the semiclassical limit of large
quantum numbers.
\end{abstract}
\section{Introduction}

Among the many problems which attend the canonical quantization of
general relativity is the fact that all gauge-invariant quantities,
which might qualify as observables, necessarily commute with the
Hamiltonian, and are therefore constants of the motion.  This
intriguing problem has received wide attention, which has been
comprehensively reviewed, for example, in \cite{Isham,Kuchar}
(see also \cite{Unruh&Wald,Au}).  It seems possible to
obtain some insight into the ``problem of time'' from simple
quantum-mechanical models which share with general relativity the
essential feature of time-reparametrization invariance, and the
purpose of the present work is to explore in some detail the
interpretation of time evolution in one such model.

It is widely thought that the time evolution of physical quantities
which is apparent to actual observers in our universe should be
understood, in some sense, as evolution relative to the state of a
physical clock, which provides an observer's local definition of time.
A particular implementation of this idea has been proposed by Rovelli
\cite{Rovelli1,Rovelli2,Rovelli3}, who defines ``evolving constants of
the motion'' by eliminating the arbitrary time coordinate $t$ in
favour of the value of a physical clock variable.  Thus, in
\cite{Rovelli1}, he considers a model of two harmonic oscillators
having the same frequency, whose equations of motion can be solved for
two functions, say $x_1(t)$ and $x_2(t)$.  Neither of these functions
gives a gauge-invariant description of time evolution, but by
eliminating $t$, Rovelli arrives at a new function $x_1(x_2)$ which is
gauge invariant (or almost so - see section 4 below) provided that
$x_2$ is regarded as a real parameter, rather than as a dynamical
variable.  Classically, at least, this defines a family of
observables, $x_1(s)$, which can be interpreted as the value of $x_1$
when $x_2 = s$.  However, the quantum-mechanical operator
corresponding to $x_1(s)$ is approximately self-adjoint (and
approximately obeys a standard Heisenberg equation of motion) only
when restricted to a subspace of the physical Hilbert space which
Rovelli calls the ``Schr\"{o}dinger regime".  Since this is a
large-quantum-number regime, he suggests that the notion of time
evolution does not exist at a fundamental level, and emerges only as a
semiclassical property of macroscopic objects.

In this paper, we study a model whose dynamical variables are an
oscillator and a free particle, the latter serving as a clock whose
reading is (classically) a measure of geometrical proper time.  The
classical and quantum dynamics of this model are discussed in Section
II where, in order to obtain a well-defined quantum theory, the free
particle is treated as the low-frequency limit of a second oscillator.
We find it possible to define a parametrized family of observables
$x(\tau)$ evolving with a parameter $\tau$ which classically coincides
with the geometrical proper time.  The linearity of the clock variable
turns out to be inessential for this purpose.  Moreover, this
evolution is governed by an exact Heisenberg equation of motion, and
is not restricted to a macroscopic regime.  The introduction of a
clock variable permits time evolution by relaxing the energy
constraint on the oscillator, and indeed the energy of the clock can
saturate this constraint.  In Section III, we investigate the
possibility of a ``test clock'' by restricting the state of the clock
to a narrow range, which approximately reproduces the constraint on
the oscillator alone.  This restriction imposes a limit on the
resolution with which the value of $x(\tau)$ can be determined, but
there is a semiclassical limit in which both the energy of the clock
and the eigenvalues of $x(\tau)$ are fairly sharply defined.  The
interpretation of time is discussed in detail in Section IV where, in
particular, we point out that the clock cannot itself be regarded as a
physically observed object, but should rather be taken as representing
the observer's intrinsic sense of time.

\section{Classical and quantum dynamics of an oscillator and clock}

Consider the Lagrangian
\begin{equation}
L = {1\over{2\lambda}}\left(\dot{x}^2 + \dot{q}^2\right) - {\lambda
\over 2}\left(\omega^2x^2 - 2E_0\right)\ ,
                                               \label{clocklagrangian}
\end{equation}
defined on a $(0+1)$-dimensional manifold with ``time'' coordinate
$t$.  The only geometrical variable is lapse function $\lambda(t)$,
which we take to be strictly positive. The action $\int L{\rm d}t$ is
easily seen to be invariant under a change of time coordinate, with
$x'(t') = x(t)$, $q'(t')=q(t)$ and $\lambda'(t')=({\rm d}t/{\rm d}t')
\lambda(t)$, where $t'$ is any increasing function of $t$. The
equations of motion
\begin{eqnarray}
{1\over\lambda}{\partial\over{\partial t}}\left({1\over\lambda}
{{\partial x}\over{\partial t}}\right)&=&-\omega^2x    \label{eofm1}\\
{1\over\lambda}{\partial\over{\partial t}}\left({1\over\lambda}
{{\partial q}\over{\partial t}}\right)&=&0\ ,            \label{eofm2}
\end{eqnarray}
obtained by varying $x$ and $q$ have the general solution
\begin{eqnarray}
x(t)&=&a_1\cos\psi(t) + a_2\sin\psi(t)              \label{gensoln1}\\
q(t)&=&c_1 + c_2\psi(t)\ ,                            \label{gensoln2}
\end{eqnarray}
so that the clock variable $q$ is linear in the proper time
$\omega^{-1}\psi(t) = \int_0^t\lambda(t'){\rm d}t'$.  The constraint
\begin{equation}
\left({{\dot{x}}\over\lambda}\right)^2 + \left({{\dot{q}}\over\lambda}
\right)^2 + \omega^2x^2 = 2E_0                     \label{constraint1}
\end{equation}
obtained by varying $\lambda$ implies the relation
\begin{equation}
{\cal C}\equiv a_1^2 + a_2^2 + c_2^2 = \rho_0^2\ , \label{constraint2}
\end{equation}
where $\rho_0^2 = 2E_0/\omega^2$.

For this model, the solution space (ignoring the factor $\Lambda$ of
gauge functions $\lambda(t)$, all of whose points are gauge
equivalent) is $S^2\times R$. With the parametrization
\begin{eqnarray}
a_1&=& \rho_0\cos\chi\cos\gamma \label{chigammaparam1} \\
a_2&=& -\rho_0\cos\chi\sin\gamma \label{chigammaparam2}\\
c_2&=& \rho_0\sin\chi\ , \label{chigammaparam3}
\end{eqnarray}
the angles $\chi$ and $\gamma$ ($-\pi/2\le\chi\le\pi/2$, $0\le\gamma<2
\pi$) provide coordinates on $S^2$, while $c_1$ is the coordinate on
$R$. The general solutions (\ref{gensoln1}) and (\ref{gensoln2}) can
be written as
\begin{eqnarray}
x(t)&=&\rho_0\cos\chi\cos\left(\psi(t)+\gamma\right)\label{gensoln3}\\
q(t)&=&c_1 +\rho_0\sin\chi\psi(t) \label{gensoln4}
\end{eqnarray}
and the presymplectic form on the solution space is
\begin{eqnarray}
\Omega&=&\omega({\rm d} a_1\wedge{\rm d} a_2 +{\rm d} c_1\wedge{\rm d}
c_2) \nonumber\\
&=&{1\over 2} \omega \rho_0^2\sin (2\chi){\rm d} \chi\wedge{\rm d}
\gamma -\omega\rho_0\cos\chi{\rm d} \chi \wedge{\rm d} c_1\ .
\label{Omega}
\end{eqnarray}

As can be seen from (\ref{gensoln3}) and (\ref{gensoln4}), time
evolution is generated by the Hamiltonian vector field
\begin{equation}
X_{\cal C} = {\partial\over{\partial\gamma}} + \rho_0\sin\chi
{\partial\over{\partial c_1}}                               \label{xc}
\end{equation}
associated with the constraint ${\cal C}$. As expected, we find that
$X_{\cal C}$ annihilates $\Omega$, $i_{X_{\cal C}}\Omega = 0$, so the
reduced phase space is found by identifying all points along each
gauge orbit, which is an integral curve of $X_{\cal C}$. Equivalently,
points of the reduced phase space correspond to distinct gauge orbits.
The angle $\chi$ is constant along each gauge orbit. For each value of
$\chi$ (except $\chi = 0, \pm\pi/2$), the orbits are helical curves on
the cylinder $S^1\times R$ coordinatized by $\gamma$ and $c_1$, with a
period along $R$ of $2\pi\rho_0\sin\chi$. Distinct orbits can
therefore be labelled by the values of $c_1$ in an interval of length
$2\pi\rho_0\sin\chi$, the end points of which are identified, since
they belong to the same orbit. The set of distinct orbits is thus
topologically $S^1$. For $\chi = 0$, the gauge orbits are circles on
the cylinder, and there is a distinct orbit for each $c_1 \in R$. For
$\chi = \pm\pi/2$, the amplitude of oscillation in (\ref{gensoln3})
vanishes, so all values of $\gamma$ are equivalent. In this case, the
$S^1$ of gauge orbits collapses to a point.

We see that the reduced phase space is the union of three sets:
\begin{enumerate}
\item[(i)]
the open disk $D_+$, which is the direct product of the interval
$0<\chi\le\pi/2$ with the circle $S^1$, the circle at $\chi=\pi/2$
being contracted to a point.  Physically, states in this region are
those in which the clock runs forwards.  The gauge-invariant
observables which distinguish the states are $\chi$, which sets the
amplitude of the $x$ oscillation and also (via the constraint) the
rate of the clock, and $c_1$, which characterises the phase of
oscillation at which the clock reads zero.
\item[(ii)]
the open disk $D_-$, which is similar to $D_+$, but corresponds to the
interval $-\pi/2\le\chi<0$ and contains those states in which the
clock runs backwards.
\item[(iii)]
the line $R_0$, with $c_1\in (-\infty,+\infty)$ and $\chi=0$. In this
region, the oscillation has its maximum amplitude, while the clock has
a constant reading $c_1$, which is the gauge invariant observable
distinguishing the states.
\end{enumerate}

This reduced phase space is {\it not} a manifold - a fact which
presents serious difficulties for the method of geometrical
quantization, and we have not succeeded in obtaining a fully
consistent quantum version of this model. A quantum theory of a
related model will be described below. Within the classical
theory, we can construct gauge-invariant observables which evolve with
proper time, as measured by the clock, by restricting attention to the
subspace $D_+\cup D_-$. Consider that region of the solution space
where $\chi\ne 0$ and identify the points $(\chi,\gamma, c_1)$ and
$(\chi, \gamma, c_1 + 2\pi\rho_0\sin\chi)$.  On the resulting space,
we replace the coordinates $c_1$ and $\gamma$ with
\begin{eqnarray}
c&=&{{c_1}\over{\rho_0\sin\chi}}                  \label{newcoords1}\\
\alpha&=&\left(\gamma - {{c_1}\over{\rho_0\sin\chi}}\right)
\hbox{\rm\ \ mod\ }2\pi\ .                          \label{newcoords2}
\end{eqnarray}
In terms of these coordinates, the generator (\ref{xc}) of gauge
orbits is
\begin{equation}
X_{\cal C} = {\partial\over{\partial c}}                 \label{newxc}
\end{equation}
and the presymplectic form (\ref{Omega}) becomes
\begin{equation}
\Omega = {1\over 2}\omega\rho_0^2\sin (2\chi){\rm d} \chi\wedge{\rm d}
\alpha\ . \label{newOmega}
\end{equation}
As expected, this $\Omega$ is independent of $c$, and is also the
symplectic form on the region $D_+\cup D_-$ of the reduced phase
space, now coordinatized by $\chi$ and $\alpha$.

With the coordinates $(\chi, \alpha, c)$, the solutions to the
equations of motion read
\begin{eqnarray}
x(t)&=&\rho_0\cos\chi\cos\left(\psi(t)+\alpha + c\right)
                                                    \label{gensoln5}\\
q(t)&=&\rho_0\sin\chi\left(\psi(t) + c\right)\ .      \label{gensoln6}
\end{eqnarray}
The quantity $\tau = \left(\psi(t) + c\right)/\omega$ can be
identified as the proper time which has elapsed since the clock read
zero and (with an obvious economy of notation) the value of $x$ at
this time is given by
\begin{equation}
x(\tau) = \rho_0\cos\chi\cos(\omega\tau + \alpha) = X\cos(\omega\tau)
+\omega^{-1}P\sin(\omega\tau)\ ,                        \label{xoftau}
\end{equation}
where
\begin{eqnarray}
X&=&\rho_0\cos\chi\cos\alpha                               \label{X}\\
P&=&-\omega\rho_0\cos\chi\sin\alpha\ .                       \label{P}
\end{eqnarray}
Since they do not depend on $c$, the quantities $X$ and $P$ are gauge
invariant.  Thus, if $\tau$ is regarded as a parameter, rather than as
standing for the expression $\left(\psi(t) + c\right)/\omega$, then
$x(\tau)$ is a gauge-invariant function of $\tau$.  Classically, this
function can be interpreted as ``the value of $x$ at proper time
$\tau$'', given that $\tau = 0$ is located at the event $q = 0$.  It
is similar to the ``evolving constants of the motion'' defined by
Rovelli \cite{Rovelli1,Rovelli2} but with an important difference:
whereas Rovelli's evolution is with respect to the actual value of a
physical clock variable, $\tau$ refers (at the classical level) to the
geometrical proper time. The symplectic form can be expressed in terms
of $X$ and $P$, with the satisfactory result
\begin{equation}
\Omega = {\rm d}X\wedge{\rm d}P\ ,                     \label{OmegaXP}
\end{equation}
so $X$ and $P$ have the canonical Poisson bracket algebra. Moreover,
the evolution of $x(\tau)$ is easily seen to be governed by the usual
equation of motion
\begin{equation}
{{{\rm d} x(\tau)}\over{{\rm d} \tau}} = \left\{x(\tau), H_0\right\}\
, \label{dxdtau}
\end{equation}
where
\begin{equation}
H_0 = {1\over 2} (P^2 + \omega^2 X^2) = {1\over 2} \left(\left({{{\rm
d} x(\tau)}\over {{\rm d} \tau}}\right)^2 + \omega^2 x(\tau)^2\right)
\label{H_0}
\end{equation}
is the Hamiltonian for the oscillator alone.

To obtain a quantum version of this model in a controlled manner, we
consider the Lagrangian
\begin{equation}
L = {1\over{2\lambda}}\left(\dot{x}^2 + \dot{q}^2\right) - {\lambda
\over 2}\left(\omega^2x^2 +N^{-2}\omega^2q^2 - 2E_0\right)\ ,
                                              \label{clocklagrangian2}
\end{equation}
where $N$ is an integer, with a view to recovering the model (\ref
{clocklagrangian}) in the limit $N\rightarrow\infty$. Somewhat
surprisingly, the results are essentially independent of $N$.  The
classical analysis is similar to that given above.  If the solutions
to the equations of motion are written as
\begin{eqnarray}
x(t)&=&a_1\cos\psi(t) + a_2\sin\psi(t)              \label{gensoln7}\\
q(t)&=&b_1\cos\left(N^{-1}\psi(t)\right)
+ b_2\sin\left(N^{-1}\psi(t)\right)\ ,                \label{gensoln8}
\end{eqnarray}
then the quantities
\begin{eqnarray}
a&=&\sqrt{{\omega\over 2}}(a_1 + ia_2)                     \label{a}\\
b&=&\sqrt{{\omega\over{2N}}}(b_2 - ib_1)                     \label{b}
\end{eqnarray}
have Poisson brackets $\{a,a^*\}=\{b,b^*\} = -i$, which will shortly
be promoted to commutators. The constraint $a_1^2 + a_2^2 +
N^{-2}(b_1^2 + b_2^2) = \rho_0^2$ can be solved in terms of the
coordinates $\chi$, $\alpha$ and $c$, replacing (\ref{gensoln5}) and
(\ref{gensoln6}) with
\begin{eqnarray}
x(t)&=&\rho_0\cos\chi\cos\left(\psi(t)+\alpha + c\right)
                                                    \label{gensoln9}\\
q(t)&=&N\rho_0\sin\chi\sin\left(N^{-1}(\psi(t) + c)\right)\ .
                                                     \label{gensoln10}
\end{eqnarray}

For any finite value of $N$, the reduced phase space of this model is
$S^2$, the angles $\chi$ and $\alpha$ ($0\le\chi\le\pi/2$,
$0\le\alpha\le 2\pi$) providing coordinates such that the points
$\chi=0,\pi/2$ are opposite poles. In particular, for $\chi = 0$, the
angle $\alpha$ in (\ref{gensoln9}) is indistinguishable from $c$, so
all values of $\alpha$ are gauge equivalent. Each point of this phase
space, of course, corresponds to a periodic oscillation, for which the
values $\psi(t) = 0, 2N\pi$ may be identified. The limit $N\rightarrow
\infty$ is actually singular, in the following sense. To achieve a
finite limit in (\ref{gensoln10}), with $\chi \ne 0$, we may assume
that $(\psi(t) + c)$ either has a finite value, or differs by a finite
amount from $N\pi$. In the limit, therefore, each point of the phase
space $S^2$ represents two distinct solutions, corresponding to the
half-periods of oscillation in which $q$ is an increasing or a
decreasing function of $\psi(t)$. Clearly, the two limits of
(\ref{gensoln10}) differ by a sign, which can be regarded as the sign
of $\chi$, and correspond to the two regions $D_{\pm}$ of the phase
space described above. To reproduce the region $R_0$, we must first
take $N\rightarrow\infty$ with $\chi\ne 0$ and then take
$\chi\rightarrow 0$, with $c$ given by (\ref{newcoords1}).

For finite values of $N$, the reduced phase space $S^2$ is a manifold,
and the model can be quantized straightforwardly. However, since this
manifold includes the point $\chi = 0$, we may anticipate some
difficulty in interpreting the variables $X$ and $P$ (equations
(\ref{X}) and (\ref{P})) as operators on the physical Hilbert space.
Since all values of $\alpha$ are gauge equivalent at $\chi = 0$, $X$
and $P$ (and hence also $x(\tau)$) are not gauge-invariant functions
on the reduced phase space, although they {\it are} gauge-invariant
functions on any region which excludes the point $\chi = 0$. In any
such region, the expressions (\ref{newOmega}) and (\ref{OmegaXP}) for
the symplectic form are still valid, so $X$ and $P$ still have the
canonical Poisson algebra.

Models of this kind can be quantized in several more or less
equivalent ways (see, for example
\cite{Rovelli1,Plyushchay,Louko,Ashtekar&Tate}). The essential result
is that one can identify gauge-invariant functions $(s_1, s_2, s_3) =
(\omega \rho_0^2/4) (\sin(2\chi)\sin\alpha, \cos(2\chi),
\sin(2\chi)\cos\alpha) $, on the reduced phase space, whose Poisson
algebra, $\{s_i, s_j\} = \epsilon_{ijk}s_k$ is the Lie algebra of
$SU(2)$, and the Casimir invariant $s_1^2+s_2^2+s_3^2$ is proportional
to the constraint ${\cal C}$. The physical Hilbert space is therefore
an irreducible representation of $SU(2)$ whose dimension is set (up to
operator ordering ambiguities) by the value of $E_0$. For our
purposes, it is convenient to construct this Hilbert space according
to the Dirac prescription, taking $a$ and $b$ as the basic variables
(a similar route is followed in \cite{Ashtekar&Tate}).

On promoting $a$ and $b$ to quantum ladder operators, with the usual
commutators $[a, a^{\dag}] = [b, b^{\dag}] = 1$, $[a, b] = [a,
b^{\dag}] = 0$, we obtain an unconstrained Hilbert space, spanned by
the vectors $|m, n\rangle$, with
\begin{equation}
a|m, n\rangle = \sqrt{m}|m-1, n\rangle\ , \ \ a^{\dag}|m, n\rangle =
\sqrt{m+1}| m+1,n\rangle
\end{equation}
\begin{equation}
b|m, n\rangle = \sqrt{n}|m, n-1\rangle\ ,\ \ b^{\dag}|m, n\rangle
= \sqrt{n+1}| m, n+1\rangle\ .
\end{equation}
The physical Hilbert space is the subspace of vectors satisfying the
constraint
\begin{equation}
(a^{\dag}a + N^{-1}b^{\dag}b)|\psi\rangle = \bar{\nu}|\psi\rangle\ ,
                                                 \label{constraint3}
\end{equation}
where $\bar{\nu}= \omega\rho_0^2/2 = E_0/\omega$. Here it is assumed
that any constant arising from factor ordering in the constraint
operator has been absorbed into $E_0$.  Quantization clearly requires
that $N\bar{\nu}$ be an integer, and we define
\begin{equation}
\bar{\nu} = \nu + \nu'/N\ ,
\end{equation}
where $\nu$ and $\nu'$ are integers, with $0\le\nu'<N$. This physical
Hilbert space is spanned by the vectors
\begin{equation}
|m\rangle\rangle = |m, N(\nu - m)+\nu'\rangle\ \ \ ,\ 0\le m\le \nu\ .
\end{equation}

We would now like to realize $X$ and $P$ as gauge-invariant operators,
acting in the physical Hilbert space, but, as anticipated, this is not
quite straightforward. Classically, we can define variables $A$ and
$A^*$, with Poisson bracket $\{A,A^*\} = -i$, by
\begin{equation}
A = \sqrt{\omega\over 2}\left(X + {i\over\omega}P\right) = a\left(
{{b^*}\over{\sqrt{b^*b}}}\right)^N\ .
\end{equation}
Quantum-mechanically, the operator ordering
\begin{eqnarray}
A&=&a\left((b^{\dag}b)^{-1/2}b^{\dag}\right)^N\\
A^{\dag}&=&a^{\dag}\left(b(b^{\dag}b)^{-1/2}\right)^N
\end{eqnarray}
ensures that these operators have the expected properties
$A|m\rangle\rangle= \sqrt{m}|m-1\rangle\rangle$ and
$A^{\dag}|m\rangle\rangle=\sqrt{m+1}|m+1\rangle\rangle$, {\it except}
that the action of $A^{\dag}$ on the maximal state
$|\nu\rangle\rangle$ is not well defined. To proceed, we introduce the
regularized operators
\begin{eqnarray}
A_{\epsilon}&=&a\left((b^{\dag}b+\epsilon)^{-1/2}b^{\dag}\right)^N
\label{Areg}\\
A^{\dag}_{\epsilon}&=&a^{\dag}\left(b(b^{\dag}b + \epsilon)^{-1/2}
\right)^N\ .\label{Adagreg}
\end{eqnarray}
These are well-defined, gauge invariant operators, and we find in
particular that $A^{\dag}_{\epsilon} |\nu\rangle\rangle=0$. We can now
define operators $A$ and $A^{\dag}$ by
\begin{eqnarray}
A|m\rangle\rangle &=&\lim_{\epsilon\to0}A_{\epsilon}|m\rangle =
\sqrt{m}|m-1\rangle\rangle \label{Aonm}\\
A^{\dag}|m\rangle\rangle
&=&\lim_{\epsilon\to0}A^{\dag}_{\epsilon}|m\rangle =\sqrt{m+1}(1 -
\delta_{m,\nu})|m+1\rangle\rangle \label{Adagonm}
\end{eqnarray}
These operators are also well defined and gauge invariant, but they
have the anomalous commutator
\begin{equation}
[A,A^{\dag}] = 1 - \theta\ ,
\end{equation}
where $\theta$ projects onto the maximal state:
$\theta|m\rangle\rangle=(1+\nu) \delta_{m,\nu}|m\rangle\rangle$. The
operators
\begin{eqnarray}
X&=&\left({1\over{2\omega}}\right)^{1/2}\left(A+A^{\dag}\right)
                                                        \label{XofA}\\
P&=&-i\left({\omega\over2}\right)^{1/2} \left(A-A^{\dag}\right)
                                                          \label{PofA}
\end{eqnarray}
are well defined, gauge invariant and self-adjoint, but they have the
anomalous commutator $[X,P] = i[A,A^{\dag}] = i(1-\theta)$. The
anomalous term is nonzero only when acting on the maximal state
$|\nu\rangle\rangle$, where the clock has the smallest energy allowed
by the constraint (\ref{constraint3}). Classically, the interpretation
of $X$ and $P$ as the position and momentum of the oscillator ``at
time $\tau = 0$'' is ambiguous in the state $\chi = 0$, where the
clock permanently reads zero, and therefore does not distinguish this
instant of time, and the anomaly can be understood as reflecting this
fact in the quantum theory. Formally, the classical versions of
(\ref{Areg}) and (\ref{Adagreg}) correspond to regularized variables
\begin{equation}
X_{\epsilon} = \rho_0\cos\chi\cos\alpha\left[1+{{2\epsilon}\over{N
\omega\rho_0^2\sin^2\chi}}\right]^{-N/2}
\end{equation}
and $P_{\epsilon} = -\omega\rho_0\tan\alpha X_{\epsilon}$. These are
truly gauge-invariant quantities, having the unique values
$X_{\epsilon} = P_{\epsilon} = 0$ at $\chi = 0$, but do not have the
canonical Poisson algebra. However, they differ significantly from the
original $X$ and $P$ only where $\sin^2\chi$ is not much greater than
$\epsilon/\omega\rho_0^2$, and when $\epsilon$ is sufficiently small,
this is a small neighbourhood of the point $\chi = 0$.

It is simple to show that
\begin{eqnarray}
\theta A&=&A^{\dag}\theta = 0\\
{[}A,A^{\dag}A{]}&=&A\\
{[}A^{\dag},A^{\dag}A{]}&=&-A^{\dag}\ ,
\end{eqnarray}
and it follows from the latter relations that the Heisenberg equation
of motion
\begin{equation}
{{{\rm d} x(\tau)}\over{{\rm d} \tau}} = i[H_0,x(\tau){]}
\label{Heisenberg}
\end{equation}
reproduces (\ref{xoftau}) in the expected way, provided that the
Hamiltonian has the factor ordering $H_0 = \omega A^{\dag}A$.

At this point, we have a formalism which, taken at face value, is
equivalent to ordinary time-dependent quantum mechanics. There is a
physical Hilbert space, spanned by the vectors $|m\rangle\rangle$,
a set of
operators $(X,P)$ on this space and a Hamiltonian $H_0$ which
generates time evolution through the standard equation of motion
(\ref{Heisenberg}). Whether this formalism should be taken at face
value is, of course, another matter, and the interpretation of the
model will be discussed in Section IV. Apparently, one can ask, and
answer, questions such as ``given that $x$ was determined to have the
value $x_1$ at time $\tau_1$, what is the probability that it has the
value $x_2$ at time $\tau_2$?''. Acording to standard quantum
mechanics, this question is legitimate only if $x_1$ and $x_2$ are
eigenvalues of $x(\tau_1)$ and $x(\tau_2)$. Because the classical
range of $x$ is restricted by the constraint and, correspondingly, the
physical Hilbert space is finite- dimensional, these eigenvalues form
a finite, discrete spectrum, which is easily found. Using
(\ref{XofA}), we can express $x(\tau)$ as $x(\tau) =
(2\omega)^{-1/2}(Ae^{-i\omega\tau} + A^{\dag}e^{i\omega\tau})$. The
properties (\ref{Aonm}) and (\ref {Adagonm}) then imply that
\begin{equation}
|x_j,\tau\rangle\rangle = N_j
\sum_{m=0}^{\nu}(2^mm!)^{-1/2}H_m(\omega^{1/2}x_j) {\rm
e}^{im\omega\tau}|m \rangle\rangle \label{xeigenstates}
\end{equation}
is an eigenvector of $x(\tau)$ with eigenvalue $x_j$, where $H_m$ is
the Hermite polynomial, $N_j$ is a normalizing constant and, in order
to truncate the series at $m = \nu$, $\omega^{1/2}x_j$ must be a zero
of $H_{\nu + 1}$. Since there are $\nu + 1$ of these zeros, the states
(\ref{xeigenstates}) span the $(\nu+1)$-dimensional Hilbert space.
Also, since $p(\tau) = \dot{x}(\tau) = \omega x(\tau + \pi/2\omega)$,
the eigenvalues of $p(\tau)$ are $\omega x_j$.

The quantum dynamics of this model can be illustrated by considering
the coherent state
\begin{equation}
|z\rangle\rangle = C(z)e^{zA^{\dag}}|0\rangle\rangle =
C(z)\sum_{m=0}^{\nu}{{z^m}\over{\sqrt{m!}}}|m\rangle\rangle\ ,
                                                 \label{coherentstate}
\end{equation}
where $C(z)$ is a normalizing factor.  The time-dependent wavefunction
for this state is
\begin{eqnarray}
\psi(x_j,\tau;z)&=&\langle\langle x_j,\tau|z\rangle\rangle \nonumber\\
&=& N_jC(z)\sum_{m=0}^{\nu}
{1\over{m!}}H_m(\omega^{1/2}x_j)\left({{z(\tau)}\over{\sqrt{2}}}
\right)^m\ , \nonumber\\
&&                                 \label{coherentwavefunction}
\end{eqnarray}
where $z(\tau) = e^{-i\omega\tau}z$. If $z = \sqrt{\omega/2}(x_0 +
(i/\omega)p_0)$, then $z(\tau) = \sqrt{\omega/2}(x_0(\tau)+
(i/\omega)p_0({\tau}))$, where $(x_0(\tau),p_0(\tau))$ is the
classical trajectory passing through $(x_0,p_0)$. For an unconstrained
oscillator (corresponding to the limit $\nu = \infty$), the coherent
state is, of course, a Gaussian wave packet, with $|\psi|^2\propto
\exp(-\omega(x- \sqrt{2/\omega}\Re z)^2)$, whose peak follows the
classical trajectory. The wavefunction (\ref{coherentwavefunction}) is
defined only at the discrete values $x_j$, and is not simply a
function of $x_j - z(\tau)$. Nevertheless, if $x_0$ is well within the
range defined by the largest and smallest $x_j$ then, even for quite
small values of $\nu$, these discrete values follow the Gaussian
packet rather closely, as illustrated in figure 1.

\section{Test clocks}

In classical general relativity, one can assess the physical
characteristics of a spacetime by examining the trajectories of ``test
particles'' which follow time-like geodes\-ics, but do not contribute
to the stress tensor. Here we consider the possibility of a ``test
clock'', which might be used to reveal the time evolution of a quantum
universe, without itself contributing significantly to the quantum
dynamics. Whether this notion is ultimately meaningful or useful
depends on how the quantum theory is to be interpreted, and this is
discussed in the following section.

Suppose, specifically, that the clock variable $q$ is deleted from the
Lagrangian (\ref{clocklagrangian}) or (\ref{clocklagrangian2}).
Classically, there seems to be a sense in which the state of the
remaining oscillator evolves with proper time as $x(\tau) =
\sqrt{2E_0/\omega^2}\cos(\omega \tau)$, although $x(\tau)$ cannot be
regarded as a gauge-independent observable.  In the quantum theory,
the Hilbert space contains only a single state, from which nothing
corresponding to this apparent time evolution can be extracted.  By
adding the clock, we are able to obtain a genuine observable $x(\tau)$
and a multi-dimensional Hilbert space on which proper time evolution
can be represented in a gauge-invariant manner.  However, the energy
constraint which applied to the original oscillator is relaxed by the
presence of the clock, which can itself saturate the constraint.  The
idea now is to see whether the apparent time-dependence of the
clockless model can be revealed by restricting attention to those
states in which the clock has only small energies, so that the energy
constraint of the clockless model is approximately realized.  It may
be anticipated that this restriction will impose a limit on the
resolution with which values of $x(\tau)$ can be determined, and we
wish to examine the nature of this limit.

If there is to be a non-trivial time evolution, several states must
still be available.  We therefore consider a (loosely) semiclassical
situation, with $\nu$ very large, taking a band of states with
$m\sim\nu$ to be actually available for use. Our calculations will now
be approximate, and we begin by finding an approximation for the
eigenstates (\ref{xeigenstates}). It is convenient at this point to
use the Schr\"odinger picture, and deal with the eigenstates
$|x_j,0\rangle\rangle$ of $X = x(0)$. When $\nu$ is large, the $\nu +
1$ zeros of $H_{\nu+1}$ corresponding to the eigenvalues of $x(\tau)$
lie roughly between $-\sqrt{2\nu}$ and $\sqrt{2\nu}$ and the spacing
between them is therefore of order $\nu^{-1/2}$. We will take these
eigenvalues to form a continuum. For large $m$, the asymptotic formula
for $H_m$ \cite{Szego},
\begin{equation}
H_m(z) \approx e^{z^2/2}{{\Gamma(m+1)}\over{\Gamma(m/2+1)}}\cos\left(
(2m+1)^{1/2}z-m\pi/2\right)
\end{equation}
yields
\widetext
\begin{equation}
\left(2^{4n+r}(4n+r)!\right)^{-1/2}H_{4n+r}(z)\approx ( 2\pi)^{-1/4}
e^{z^2/2}n^{-1/4}c_r(\sqrt{8n}z)\ ,\ \ r = 0 \cdots 3  \label{Happrox}
\end{equation}
where $c_r(\theta) = \cos(\theta - r\pi/2)$. Taking $\nu$ to be of the
form $4\nu'+3$ and $m = 4\nu'\alpha + r$ in (\ref{xeigenstates}),
replacing the sum on $\alpha$ by an integral, and defining
\begin{equation}
\hat{\xi} = \left({\omega\over{2\nu}}\right)^{1/2}X\ ,      \label{xi}
\end{equation}
we obtain eigenstates of $\hat{\xi}$ in the form
\begin{equation}
|\xi\rangle\rangle = \sqrt{\nu\over{2\pi}}\int_0^1{\rm d} \alpha\
\alpha^{-1/4} \sum_{r=0}^3c_r(2\nu\sqrt{\alpha}\xi)|\alpha,
r\rangle\rangle\ ,
\end{equation}
which are orthonormal in the limit $\nu\rightarrow \infty$.

Ladder operators $A^{(1)}\equiv A$ and $A^{(-1)}\equiv A^{\dag}$ are
realized in this aproximation by
\begin{eqnarray}
A^{(\mu)}|\xi\rangle\rangle
&=&\sqrt{\mu}\left(\xi - {\mu\over{2\nu}} {\partial
\over{\partial\xi}}\right)|\xi\rangle\rangle           \label{Aonxi}\\
A^{(\mu)}|\alpha,r\rangle\rangle
&=&\sqrt{\nu}\left(\alpha^{1/2} - {\mu\over\nu}
\alpha^{1/4}{\partial\over{\partial\alpha}}\alpha^{1/4}\right)
\nonumber\\ &&\qquad \times|\alpha,
r-\mu\rangle\rangle\ .                               \label{Aonalphar}
\end{eqnarray}
Because the maximal state annihilated by $A^{\dag}$ now has zero
measure, these operators have the canonical commutator $[A,A^{\dag}]
=1$.  For the same reason, the coherent state is now an eigenstate of
$A$.  Corresponding to the rescaled variable $\xi$ (\ref{xi}) we
define
\begin{equation}
A|z\rangle\rangle = \sqrt{\nu}z|z\rangle\rangle
\end{equation}
and obtain the usual wavefunction $\psi(\xi,z)=\langle\langle
\xi|z\rangle\rangle \sim \exp(-\nu(\xi-z)^2)$.

Supposing that the states actually available for use are those in a
narrow range of $\alpha$, we introduce a window function $f(\alpha)$
and define approximate eigenstates of $\hat{\xi}$ by
\begin{equation}
|\xi\rangle\rangle^f = \int_0^1{\rm d}\alpha\ \alpha^{-1/4}f(\alpha)
\sum_{r=0}^3 c_r(2\nu\sqrt{\alpha}\xi)|\alpha,r\rangle\rangle\ .
\label{approxxistates}
\end{equation}
In the same way, we can define an approximate coherent state
\begin{equation}
|z\rangle\rangle^f = \int_0^1{\rm d}\alpha\
\alpha^{-1/4}f(\alpha)\sum_{r=0}^3\psi_r
(\alpha,z)|\alpha,r\rangle\rangle\ ,            \label{approxcoherent}
\end{equation}
where
\begin{equation}
\psi_r(\alpha,z) = \int_{-1}^1{\rm d}\xi \psi(\xi,z)c_r(\alpha)\ .
                                                  \label{approxwavefn}
\end{equation}
As in (\ref{coherentwavefunction}) the time evolution of this state is
obtained by making the replacement $z\rightarrow z(\tau)$.

To obtain approximate analytical results, we define $\gamma =
\sqrt{\alpha}$ and choose a Gaussian window function $f(\gamma)=f_0
\exp(-(\gamma - \gamma_0)^2/2\Delta)$, where $f_0$ is the appropriate
normalization factor. We choose $\gamma_0$ to be a little less than
the maximal value $\gamma=1$, and assume that $\Delta$ is small enough
for the limits of integration to be extended to infinity.  For the
uncertainty $\Delta\xi$, defined by $(\Delta\xi)^2 =\ ^f
\langle\langle\xi| \hat{\xi}^2|\xi\rangle\rangle^f - (^f\langle\langle
\xi|\hat{\xi}|\xi\rangle\rangle^f)^2$, we find
\begin{equation}
\Delta\xi = {1\over{2\nu\sqrt{2\Delta}}}\ .
\end{equation}
We see that, for a fixed window width $\Delta$, it is possible to make
$\xi$ arbitrarily sharp when $\nu$ is sufficiently large.  On the
other hand, when $\Delta$ is made small, in effect imposing the
constraint of the clockless model, it becomes impossible to construct
sharp eigenstates of $\xi$, reflecting the fact that, without the
clock, there is no gauge-invariant operator corresponding to $X$ or
$\hat{\xi}$. Given the coherent state $|z\rangle\rangle^f$, we can
estimate the probability amplitude for finding the oscillator at the
position $\xi$ at time $\tau$. Assuming that the limits of integration
in (\ref{approxwavefn}) can be extended to infinity, which will be
valid if $\nu$ is large enough and if $|\bar{\xi}|<1$, where
$\bar{\xi} = \Re z$, we obtain
\begin{equation}
\ ^f\langle\langle \xi|z(\tau)\rangle\rangle^f
\sim\exp\left[-{{\nu^2\Delta}\over {1+\nu\Delta}}\left(\xi -
\bar{\xi}(\tau)\right)^2\right]\ ,                     \label{probxiz}
\end{equation}
where $\sim$ indicates the omission of a normalization factor and a
phase. For a fixed window width $\Delta$, this amplitude becomes
sharply peaked at the classical trajectory in the semiclassical limit
$\nu\rightarrow\infty$.  However, when $\Delta$ becomes very small,
then (i) there is no longer a sharp peak at the classical trajectory
and (ii) the probability becomes time-independent.  Qualitatively, at
least, we recover the situation in the absence of the clock, where
there is no gauge-invariant operator corresponding to $\xi$ and all
gauge-invariant amplitudes are time-independent (except possibly for
trivial phase factors).

It is apparent from these results that, as expected, we cannot fix the
energy of the clock without losing both the observable $\xi$ and the
time dependence which the clock was intended to reveal. However, it
seems that one can find situations (for example, by taking $\nu$
large, with $\Delta\nu$ fixed) in which both the energy of the clock
and the eigenstates of $\hat{\xi}$ are fairly sharply defined, and in
which the notion of a test clock is therefore meaningful.

\section{Discussion}

The absence of time-dependent observables in a
time-reparametrization-invariant system can perhaps be understood by
recognising that a complete description of a closed system is
inevitably from the point of view of an observer external to the
system. As viewed by such an observer, the system has, say, $d+1$
dimensions, of which one corresponds to the ``time'' coordinate $t$
(so the models considered in this paper have $d=0$), and the ``time''
which passes inside the system is quite unrelated to any ``time''
experienced by this observer. For this observer, a state of the
$(d+1)$-dimensional system corresponds to what an observer internal to
the system might be supposed to regard as an entire history of a
$d$-dimensional system. The observable quantities whose values
distinguish one state of the system from another therefore
characterise entire histories of the system, and cannot evolve with
time in any straightforward sense.

For a classical system which is {\it not} time-re\-par\-ametriz\-ation
invariant, this does not normally present a problem.  Consider, for
example, the model (\ref{clocklagrangian}) with $\lambda$ set equal
to 1.  The general solutions to the equations of motion are of the
form $x(t) = X\cos(\omega t) + (P/\omega)\sin(\omega t)$ and $q(t) = Q
+ \Pi t$.  There are four independent observables, $(X,P,Q,\Pi)$,
since a set of values for these quantities determines a history of the
system.  However, for a fixed value of $t$, the quantities $x(t)$ and
$q(t)$, for example, are linear combinations of $(X,P,Q,\Pi)$ and are
also observables. From such families of observables, parametrized by
$t$, we can obviously construct functions such as $x(t)$ which an
internal observer might construe as representing the evolution with
time of a single observable $x$ belonging to his 0-dimensional
``space''.  This assumes that the internal observer, who does not
appear explicitly in our model, does not significantly disturb the
objects represented by $x$ and $q$.  If suitably equipped, he will be
able to inspect two independent objects, an oscillator and a ``free
particle'', associated respectively with the pairs of phase-space
coordinates $(X,P)$ and $(Q,\Pi)$.

In the case of a time-reparametrizarion-invariant system, the families
$x(t)$ and $q(t)$ have no gauge-invariant meaning, since
they depend on $t$ through the undetermined lapse function $\lambda(t)
$. The gauge-invariant quantities which might qualify as observables
are necessarily global quantities, in the sense that they cannot be
associated with any particular value of $t$, and cannot evolve with
time in any straightforward sense. For example, the gauge-invariant
variables $\chi$ and $\alpha$ in (\ref{gensoln5}) and (\ref{gensoln6})
specify the amplitude of the $x$ oscillation and the phase of this
oscillation corresponding to $q=0$. One way out of this difficulty is
to suppose that a specific lapse function is spontaneously selected,
so that gauge invariance is broken, and gauge-variant quantities such
as $x(t)$ become physically meaningful.  In general, this strategy
seems unsound, since no physical principle gives rise to a preferred
function $\lambda(t)$.  In effect, the reparametrization invariance is
removed artificially, and the constraint obtained by varying
$\lambda(t)$ is then poorly justified.  We note, however, that recent
attempts to adapt the de Broglie-Bohm interpretation of quantum
mechanics to quantum gravity \cite{Alwis&MacIntire,Shtanov} do appear
to produce a preferred foliation of spacetime and hence a natural
definition of time evolution.

It is tempting to account for time evolution in the following way.
Consider the reparametrization-invariant model consisting just of a
single oscillator (the model (\ref{clocklagrangian}) with $q$
omitted). The solution to its equation of motion is of the form
\begin{equation}
x(t) = a\cos\left(\psi(t) + \gamma\right)\ .
\end{equation}
The amplitude $a$ is constrained to have the value $\rho_0$, so there
are no gauge-invariant variables, and the $(0+1)$-dimensional system
has just one state available to it. Nevertheless, it seems intuitively
clear that an observer inside the $(0+1)$-dimensional universe would
perceive an oscillation of the form $X(\tau) =
\rho_0\cos(\omega\tau)$, where $\tau$ is the proper time along this
observer's world line. Moreover, this assertion seems to be
gauge-independent, since the lapse function serves only to relate
$\tau$ to an arbitrary coordinate $t$, which is irrelevant to the
observer. Indeed, it is essentially this line of argument which
enables an account to be given in elementary applications of general
relativity of the time-dependent appearance of the universe to an
observer. The discrepancy between the time evolution perceived by an
internal observer and the single state apparent to an external
observer arises from the fact that our model includes no description
of the clock from which the internal observer gains his sense of time.

To reconstruct what an internal observer's experience might be, it
seems plausible that we must incorporate in our model at least a
rudimentary description of this observer's clock. Classically, the
variable $q$, whose behaviour is exhibited in (\ref{gensoln6}) mimics
an instrument whose reading is linear in the geometrical proper time
$\tau$. The quantity $x(\tau)$ defined in (\ref{xoftau}) defines a
family of gauge-invariant observables which can be construed as giving
the value of $x$ which would be perceived by an internal observer when
a proper time $\tau$ has elapsed since his clock read zero.  Moreover,
the evolution of $x(\tau)$ with $\tau$ is governed by an equation of
motion (\ref{dxdtau}) of the standard Hamiltonian form.
Quantum-mechanically, this equation of motion translates to a
Heisenberg equation (\ref{Heisenberg}), again of the standard form. It
thus seems consistent with the Copenhagen interpretation of quantum
mechanics to suppose that an external observer might determine the
state of the ``universe'' by measuring the value of $x(\tau_1)$, say
with the result $x_1$. One can then compute the probability
$|\langle\langle x_2,\tau_2|x_1,\tau_1\rangle\rangle |^2$ that a
measurement of $x(\tau_2)$ will yield the value $x_2$. This
probability may plausibly be interpreted as the probability that an
internal observer, having determined the position of the oscillator as
$x_1$ will obtain the value $x_2$ from a measurement made after an
interval $\tau_2-\tau_1$ of his proper time.

To this extent, we recover precisely the usual formulation of
time-dependent quantum mechanics. However, the status of the clock
variable in this formulation requires further thought. In addition to
$x(\tau)$, we can construct from (\ref{gensoln6}) the family of
gauge-invariant variables $q(\tau) = Q\tau$, where
$Q=\omega\rho_0\sin\chi =\omega\sqrt{\rho_0^2 - X^2 - (P/\omega)^2}$.
Classically, the values of $x(\tau)$ and $q(\tau)$ can be determined
simultaneously, and it might appear that these values represent the
results of inspecting both the oscillator and the clock at time
$\tau$. However, in contrast to the non-reparametrization-invariant
system, the physical reduced phase space is now 2-dimensional. The two
coordinates $X$ and $P$ are just sufficient to represent the position
and momentum of the oscillator, and there are no further coordinates
available to represent the clock as an independent dynamical object.
As emphasised, for example, by Unruh \cite{Unruh}, who has also
considered the two-oscillator model, the variables $a_1$ and $b_1$, in
(\ref{gensoln7}) and (\ref{gensoln8}), which one would ordinarily want
to treat as Schr\"{o}dinger picture operators representing the
oscillator and clock positions, do not correspond to operators on the
physical Hilbert space. Indeed, the variable $x(\tau)$ in
(\ref{xoftau})can be defined only when we solve the constraint,
eliminating $q$ as an independent dynamical variable. Correspondingly,
the evolution of $q(\tau)$ cannot be obtained from a Hamiltonian
equation of motion, since $Q$ commutes with $H_0$ (equation
(\ref{H_0})). Quantum-mechanically, $x(\tau)$ and $q(\tau)$ do not
commute, so their values cannot be determined simultaneously.
Moreover, the Hilbert space of this model is finite-dimensional, so
the readings of any object we wish to regard as a clock must be drawn
from a finite, discrete set of eigenvalues. From these considerations,
it is apparent that a statement such as $x(\tau_1) = x_1$ cannot be
taken as implying that both the oscillator and the clock have been
inspected, yielding the value $x_1$ for the position of the oscillator
and the value $\tau_1$ for the time.

Thus, $\tau$ cannot be regarded as a phenomenological time
deduced by an observer from his inspection of a physical clock which
reads $q$. Rather, the definition of $x(\tau)$ reflects a decision on
the part of an external observer (or of a theoretician) to study time
evolution from a particular point of view. One might, for example,
imagine an automatic observing apparatus, which performs a sequence of
tasks under the control of an internal clock, $q$. To study time
evolution ``from the point of view'' of this apparatus, it is natural
to solve the Hamiltonian constraint by eliminating the unobserved
quantity $q$, in order to arrive at the time-dependent observable
$x(\tau)$. In this sense, $\tau$ represents a ``Heraclitian'' time of
the kind sought by Unruh and Wald \cite{Unruh&Wald}, which ``sets the
conditions'' for a measurement to be made.  Like the parameter $t$ in
a non-reparametrization-invariant system, $\tau$ itself is not a
measurable quantity.

Furthermore, it appears quite unnecessary for the eliminated variable
to be linear, or even monotonic, in $\tau$. The physical Hilbert space
associated with the model (\ref{clocklagrangian2}) and the algebra of
the gauge-invariant operators $X$ and $P$ obtained by eliminating $q$
is independent of the parameter $N$ which determines the frequency of
the $q$ oscillator. The role of the clock $q$ in passing from either
(\ref{gensoln5}) and (\ref{gensoln6}) or (\ref{gensoln9}) and
(\ref{gensoln10}) to the gauge-invariant observable (\ref{xoftau}) is
that it distinguishes a history in which, for example,
$x=\rho_0\cos\chi\cos\alpha$ when $q=0$ from one in which some other
value of $q$ corresponds to this value of $x$. In essence, the
definition of a gauge-invariant observable $x(\tau)$ becomes possible
when an origin $\tau = 0$ can be specified in a coordinate-independent
manner, and the role of $q$ is to provide this origin.

The view of time evolution described here is similar, though not
identical, to that proposed by Rovelli
\cite{Rovelli1,Rovelli2,Rovelli3}. In \cite{Rovelli1}, Rovelli
considers essentially the model (\ref{clocklagrangian2}) with $N=1$.
He defines a gauge-invariant classical observable given, in our
notation, by
\begin{equation}
x(s) = \rho_0\cos\chi\cos\left[\alpha + \cos^{-1}\left({s\over{\rho_0
\sin\chi}}\right)\right]\ ,                          \label{Rovellisx}
\end{equation}
which is the value of $x$ when $q=s$. Like our $\tau$, the clock time
$s$ can assume a continuous range of real values. (Note also that,
like our $x(\tau)$, this $x(s)$ is gauge-variant at $\chi = 0$, since
it depends on $\alpha$, though in fact only $x(s=0)$ is well-defined.)
The qualitative discussion given in \cite{Rovelli2} suggests that
Rovelli wishes to regard $x(s)$ as representing the evolution of $x$
relative to the time recorded by a physical clock. For the reasons
given above, however, it is not possible to regard $s$ and $x(s)$ as
the two results of inspecting a clock and an oscillator
simultaneously.

The operator corresponding to (\ref{Rovellisx}) is approximately
self-adjoint, and obeys an approximate Heisenberg equation of motion,
only when restricted to a region of the physical Hilbert space which
Rovelli calls the ``Schr\"odinger regime'', corresponding roughly to
states in which the wavefunction $\psi(x,q)$ is sharply peaked around
the classical trajectory.  For this reason, Rovelli contends that the
notion of time does not exist at a fundamental level, but emerges only
in a suitable semiclassical limit. By contrast, our operator $x(\tau)$
is exactly self-adjoint, and obeys the exact Heisenberg equation
(\ref{Heisenberg}), with respect to the variable $\tau$ which
classically corresponds to the geometrical proper time, and this
provides a counterexample to Rovelli's contention.

We would like, of course, to speculate that the view of time evolution
proposed here can be extended to canonically quantized general
relativity. In such an extension, it seems likely that the role of
$\tau$ would be played by the proper time along the trajectory of an
observer for whose observations we wish to account, while $x(\tau)$
would correspond to local observables defined on this trajectory. To
define, say, time-dependent observables throughout a space-like
hypersurface, one would presumably need to introduce a space-filling
family of observers (or, at least, their clocks). These might
correspond to a reference fluid of the kind described by Brown and
Kucha\v{r} \cite{Brown&Kuchar}, though the relationship of our
interpretation to that proposed by these authors is not entirely clear
to us.

\acknowledgments

We are grateful to Gabor Kunstatter for enlightening discussions, and
to Chris Isham and Steve Carlip for comments on a draft of this paper.
IDL would like to thank the Winnipeg Institute for Theoretical
Physics, where this work began, for its hospitality and financial
support.  RJE would like to acknowledge the financial support of the
Natural Sciences and Engineering Research Council of Canada.

\newpage
{\bf Figure caption}

FIG.\ 1  The squared magnitude of the wavefunction
(\ref{coherentwavefunction})(circles) for a coherent state of
the constrained oscillator, which is defined only at the discrete
values $x_j$, compared with the corresponding Gaussian wavepacket
(solid curve) for the standard unconstrained oscillator. The
normalization of the Gaussian packet has been adjusted so that the
peaks of both wavefunctions have the same height. Time evolution is
depicted over a half cycle of oscillation, with amplitude slightly
smaller than the largest eigenvalue $x_j$. In this case, $\nu = 10$.


\newpage
\begin{references}
%
\bibitem{Isham}
C. Isham, {\it Canonical quantum gravity and the problem of time}
Lectures presented at the NATO Advanced Study Institute, Salamanca,
June 1992
%
\bibitem{Kuchar}
K. V. Kucha\v{r} in {\it Proceedings of the 4th Canadian Conference on
General Relativity and Relativistic Astrophysics} ed. G. Kunstatter,
D. E. Vincent and J. G. Williams (World Scientific, Singapore, 1992) p
211
%
\bibitem{Unruh&Wald}
W. G. Unruh and R. Wald, Phys. Rev. D {\bf 40}, 2598 (1989)
%
\bibitem{Au}
G. K. Au, {\it The quest for quantum gravity} University of Melbourne
preprint UM-P-95/24  (gr-qc/9506001) (1995)
%
\bibitem{Rovelli1}
C. Rovelli, Phys. Rev. D {\bf 42}, 2638 (1990)
%
\bibitem{Rovelli2}
C. Rovelli, Phys. Rev. D {\bf 43}, 442 (1991)
%
\bibitem{Rovelli3}
C. Rovelli, Class. Quantum Grav. {\bf 8}, 297, 317  (1991)
%
\bibitem{Plyushchay}
M. S. Plyushchay, Nucl. Phys. B {\bf 362}, 54 (1991)
%
\bibitem{Louko}
J. Louko, Phys. Rev. D {\bf 48}, 2710 (1993)
%
\bibitem{Ashtekar&Tate}
A. Ashtekar and R. S. Tate, J. Math. Phys. {\bf 35}, 6434 (1994)
%
\bibitem{Szego}
G. Szeg\"o, {\it Othogonal Polynomials} American Mathematical Society
Colloquium Publications Vol XXIII (1939)
%
\bibitem{Alwis&MacIntire}
S. P. de Alwis and D.A. MacIntire, Phys. Rev. D {\bf 50}, 5164 (1994)
%
\bibitem{Shtanov}
Y. Shtanov, {\it Pilot wave quantum gravity}, Bogolyubov Institute
preprint ITP-95-8E (gr-qc/9503005) (1995)
%
\bibitem{Unruh}
W. G. Unruh, {\it No time and quantum gravity}, talk presented at the
NATO Advanced Study Institute, Banff, 1990
%
\bibitem{Brown&Kuchar}
J. D. Brown and K. V. Kuch\v{r}, {\it Dust as a standard of space and
time in canonical quantum gravity}  North Carolina State
University/University of Utah preprint (gr-qc/9409001) (1994)

\end{references}
\end{document}